# A Transit of Venus Possibly Misinterpreted as an Unaided-Eye Sunspot Observation in China on 9 December 1874

Hisashi Hayakawa[1,2*] • Mitsuru Sôma[3] • Kiyotaka Tanikawa[3] • David M. Willis[2, 4] • Matthew N. Wild[2] • Lee T. Macdonald[5] • Shinsuke Imada[6] • Kentaro Hattori[7] • F. Richard Stephenson[8]

[1]Graduate School of Letters, Osaka University, 5600043, Toyonaka, Japan (JSPS Research Fellow).

[2]Rutherford Appleton Laboratory, Chilton, Didcot, Oxon OX11 0QX, UK

[3]National Astronomical Observatory of Japan, 1818588, Mitaka, Japan

[4]Centre for Fusion, Space and Astrophysics, Department of Physics, University of Warwick, Coventry CV4 7AL, UK

[5]History of Science Museum, Oxford OX1 3AZ, UK

[6]Institute for Space-Earth Environmental Research, Nagoya University, 4640814, Nagoya, Japan

[7]Graduate School of Science, Kyoto University, Kyoto, Kitashirakawa Oiwake-cho, Sakyo-ku, Kyoto, 6068502, Japan

[8]Department of Physics, University of Durham, Durham DH1 3LE, UK

* hayakawa@kwasan.kyoto-u.ac.jp; hisashi.hayakawa@stfc.ac.uk;
http://orcid.org/0000-0001-5370-3365

**Abstract**

Large sunspots can be observed with the unaided eye under suitable atmospheric seeing conditions. Such observations are of particular value because the frequency of their appearance provides an approximate indication of the prevailing level of solar activity. Unaided-eye sunspot observations can be traced back well before the start of telescopic observations of the Sun, especially in the East Asian historical records. It is therefore important to compare more-modern, unaided-eye sunspot observations with the results of telescopic sunspot observations, to gain a better understanding of the nature of the unaided-eye sunspot records. A previous comparison of Chinese unaided-eye sunspot records and Greenwich photo-heliographic results between 1874 and 1918 indicated that a few of the unaided-eye observations were apparently not supported by direct photographic evidence of at least one sunspot with a large area. This article reveals that one of the Chinese unaided-eye observations had possibly captured the transit of Venus on 09 December 1874. The Chinese sunspot records on this date are compared with Western sunspot observations on the same day. It is concluded that sunspots on the solar disk were quite small and the transit of Venus was probably misinterpreted as a sunspot (black spot) by the Chinese local scholars. This case indicates that sunspots or comparable "obscuring" objects with an area as large as 1000 millionths of the solar disk could easily have been seen with the unaided eye under suitable seeing conditions. It also confirms the visibility of sunspots near the solar limb with the unaided eye. This study provides an explanation of the apparent discrepancy between the Chinese unaided-eye sunspot observation on 09 December 1874 and the Western sunspot observations using telescopes, as well as a basis for further discussion on the negative pairs in 1900 and 1911, apparently without sufficiently large area.

**1. Introduction**

The sunspot number has been used as a measure of solar activity for datasets spanning around 400 years (*e.g.* Waldmeier, 1961; Hoyt and Schatten, 1998; Vaquero, 2007a; Vaquero and Vázquez, 2009; Hathaway, 2010; Owens, 2013; Clette *et al*., 2014; Clette and Lefèvre, 2016; Inceoglu *et al*., 2014; Vaquero *et al*., 2016; Svalgaard and Schatten, 2016). Detailed sunspot observations using telescopes have been preserved in the scientific literature since the introduction of the telescope, in 1610, not only in Europe (*e.g.* Vaquero, 2007a; Arlt, 2008, 2009, 2011; Usoskin *et al*., 2015; Arlt *et al*., 2016; Aparicio *et al*., 2014; Carrasco *et al*., 2015a, 2015b; Willis *et al*., 2013a, 2013b, 2016; Lefèvre *et al*., 2016; Lockwood *et al*., 2016a, 2016b, 2017; Carrasco and Vaquero, 2016; Cliver, 2017; Svalgaard, 2017) but also in North America (*e.g.* Domínguez-Castro, Gallego, and Vaquero, 2017; Denig and McVaugh, 2017) and East Asia (*e.g.* Yamamoto, 1935, 1937; Koyama, 1985; Hayakawa *et al*., 2018a, 2018b; Fujiyama *et al*., 2019).

However, even before telescopic sunspot observations, there is a considerable number of unaided-eye sunspot observations (also known as "naked-eye" sunspot observations) during historical times (*e.g.* Vaquero, 2007a; Vaquero and Vázquez, 2009). The history of unaided-eye sunspot observations can be traced back to at least 467 BCE, with descriptive records of unaided-eye sunspot observations (Vaquero, 2007a, 2007b), and back to 1128 CE for a descriptive record accompanied by a schematic sunspot drawing (Stephenson and Willis, 1999; Willis and Stephenson, 2001). In this coverage, East Asian historical documents provide the majority of unaided-eye sunspot records (*e.g.* Keimatsu, 1970; Wittman and Xu, 1987; Yau and Stephenson, 1988; Yang *et al*., 1998; Xu, Pankenier, and Jiang, 2000; Lee *et al*., 2004; Hayakawa *et al*., 2015, 2017a, 2017b, 2017c, 2017e, 2019; Tamazawa *et al*., 2017), and records from Europe and West Asia add further valuable information (Vysottsky, 1949; Goldstein, 1969; Vaquero and Gallego, 2002; Hayakawa *et al*., 2017a).

These unaided-eye sunspot records provide an important resource for studies of long-term solar activity (*e.g.* Yang *et al*., 1998; Vaquero, Gallego, and García, 2002; Lee *et al*., 2004) and extreme space-weather events (*e.g.* Willis and Stephenson, 2001; Willis *et al*., 2005; Hayakawa *et al*., 2017c) and hence contribute to a better understanding of both solar physics and solar-terrestrial relationships.

Therefore, the specific conditions under which sunspots can be seen with the unaided eye have been discussed in the literature. While Eddy (1980) regarded the unaided-eye sunspot records as direct proxies of solar activity, Willis, Easterbrook, and Stephenson (1980) examined the seasonal distribution of East Asian sunspot records, and found biases arising from a seasonal variation in the atmospheric seeing conditions. Heath (1994) made long-term unaided-eye sunspot observations from 1959 to 1993 and found a good correlation with contemporary sunspot number, despite the possible influence from weather and seeing conditions. Wade (1994) made sunspot observations with the unaided eye from 1980 to 1992 and showed that the heliographic latitude does not influence the visibility of unaided-eye sunspot observations. It is also known that sunspot area has a good correlation with sunspot number and it is at least reasonable to assume that large sunspots appear more frequently during the active phase of the solar cycle (Hathaway *et al*., 2002).

In this context, it is important to compare unaided-eye sunspot records with telescopic sunspot observations, in order to understand the visibility conditions of unaided-eye sunspot records relating to the historical epoch. Vaquero *et al*. (2004) examined an unaided-eye sunspot report in 1612 by Galileo Galilei in comparison with his contemporary sunspot drawing, to show that the area of the sunspot was roughly 2000 millionths of the visible solar hemisphere (msh). Hayakawa *et al*. (2017c) presented an unaided-eye sunspot record at Nagoya in 1770, together with contemporary sunspot drawings by Staudach (see also Arlt, 2008, 2009; Svalgaard, 2017) and showed that the area of this sunspot was up to 6000 msh. Willis *et al*. (2018) considered the case of essentially simultaneous unaided-eye sunspot observations on 10 and 11 February 1917, which were recorded in a ship's log and a Chinese local treatise, by comparing these observations with contemporary heliograms acquired at the Dehra Dun Observatory in India that show a large sunspot with an area of 3590 msh.

One of the most systematic comparisons was carried out by Willis, Davda, and Stephenson (1996), who compared ten Chinese unaided-eye sunspot observations from 1874 to 1918, based on the catalogue of Yau and Stephenson (1998), directly with photo-heliographic results published by the Royal Observatory, Greenwich since 1874 (Royal Greenwich Observatory, 1907). In that investigation, Willis, Davda and Stephenson (1996) found that five direct comparisons were positive, three were negative, one was marginal, and one was without data. Reconstructed solar images or original contact prints were presented for each Chinese observational date. Willis, Davda, and Stephenson (1996) also considered possible cases of inaccurate dating of the Chinese historical records. Allowing for possible alternative dates arising from transcription errors in Local Treatises, eight comparisons were positive and two were negative for the interval 1874 – 1918. The results of the investigation by Willis, Davda, and Stephenson (1996) have provided insights into the reliability of unaided-eye sunspot observations in history (*e.g*. Usoskin *et al*., 2015; Usoskin, 2017).

The report of an unaided-eye sunspot observation on 09 December 1874 was one of the initially "negative" cases, without the photographic support of sunspots with a sufficiently large area. Willis, Davda, and Stephenson (1996) noted the possibility of an inaccuracy in the dating in the Chinese observation, resulting from a scribal error, and suggested 12 December 1874 as the most favorable alternative day in the lunar month, when there was a relatively large sunspot group with a whole spot area of 823 millionths of the solar disk (msd) in projected area, as reconstructed from the sunspot area and sunspot position tabulated in the *Greenwich Photoheliographic Results* (see their Figure 12).

However, there is the possibility of another interpretation of the Chinese sunspot record on 09 December 1874. It is known that a transit of Venus occurred on this day and worldwide observations were carried out to measure the distance between the Earth and the Sun (*e.g*. Russell, 1892; Lu and Shi, 2007; Ratcliff, 2008; Lu and Li, 2013; Orchiston, 2004; Orchiston *et al*., 2015). Transits of Venus occur periodically with successive intervals of 105.5 years and 8 years, and 121.5 years and 8 years. It is also known that some historians in Europe and West Asia have misinterpreted sunspots as transits of Venus or Mercury (*e.g.* Adams *et al*., 1947; Goldstein, 1969). Therefore, the possibility of a misinterpretation of the transit of Venus as a sunspot in 1874 is considered by comparing the contemporary observations of both the transit of Venus and the sunspot observations, and also by considering the duration and visibility of the transit of Venus.

**2. Method and Source Documents**

In order to achieve this purpose, we have examined three Chinese sunspot records; one in each of the *Shíménxiànzhì* (石門縣志), the *Zhāngyànzhì* (張堰志), and the *Guō Sōngdào Rìjì* (郭嵩燾日記). It is the record in the *Shíménxiànzhì* that Yau and Stephenson (1988) and Willis, Davda, and Stephenson (1996) presented for 09 December 1874, whereas Lu and Li (2013) independently associated with the Venus transit. We have also examined the observational reports of the transit of Venus in China and Japan, carried out by Western astronomers. In the case of China, we have examined Chinese newspapers by Western scholars, such as the *Peking Magazine* (中西聞見錄:

hereafter, PM) and the *Shēnbào* (申報: hereafter, SB). In the case of Japan, we have examined the observational report by Francisco Diaz Covarrubias, a Mexican astronomer.

We have reconstructed the duration and visibility of the transit of Venus on 9 December 1874, based on the ephemeris data available from the Jet Propulsion Laboratory (JPL) DE430. The references to historical sources are shown in Appendix 1, with abbreviations in English and detailed references in their original language as a matter of traceability.

We then examine the contemporary sunspot observations using telescopes, to determine both the observational site and the sunspot areas. We compute the location and apparent area of Venus, regarded as a misinterpreted sunspot, and consider the visibility threshold of unaided-eye sunspot observations and the associated meteorological contribution to East Asian sunspot observations in history.

When comparing the area of the silhouette of Venus with the areas (umbral and whole-spot) of both isolated sunspots and sunspot groups, it is crucially important to use the correct units. As discussed in detail by Willis *et al.* (2013, 2016), in the case of the Greenwich Photo-heliographic Results in 1874 – 1976, the umbral and whole-spot areas of sunspots are first measured on the solar photograph with the aid of a micrometer and expressed in millionths of the solar disk (msd). Note that the "whole-spot" area in the Greenwich Photoheliographic Results means the area of the whole sunspot group as a combination of all sunspots in the group and includes not only umbral area but also penumbral area. However, as these sunspots are located on the Sun's visible surface (photosphere), the "projected" areas are corrected for foreshortening and these "corrected" areas are expressed in millionths of the visible solar hemisphere (msh). The normalization factor for projected areas is $\pi R^2$, where $R$ denotes the radius of the Sun, whereas the normalization factor for corrected areas is $2\pi R^2$. The areas of sunspot groups cited in the Introduction are correctly expressed in millionths of the visible solar hemisphere (msh) because these are the units quoted in the articles referenced.

As the silhouette of Venus appears on the solar disk, comparisons between the apparent area of Venus and the areas of sunspots must be made using projected sunspot areas, measured in msd. The "Measures Section" of the *Greenwich Photo-heliographic Results* gives corrected umbral and whole-spot daily sunspot areas for each sunspot group; the "Ledgers Section" gives both projected and corrected daily sunspot areas for each sunspot group; and the "Total Areas Section" gives the total daily projected umbral and whole-spot areas.

**3. Chinese "Sunspot" Records on 09 December 1874**

The Chinese "sunspot" records dated 09 December 1874 are found in the *Shíménxiànzhì*, the *Zhāngyànzhì*, and the *Guō Sōngdào Rìjì*, as shown below with their transcriptions and translations:

*Shíménxiànzhì*, v.11, f.22b

*Transcription*：同治十三年⋯十一月朔日中有黑子。

*Translation*: On 09 December 1874, there was a black spot in the Sun.

*Zhāngyànzhì*, v.11, f.6b

*Transcription*：同治十三年⋯十一月朔日中有黑子。

*Translation*: On 09 December 1874, there was a black spot in the Sun.

*Guō Sōngdào Rìjì*, v.2, p.845. [*Guō Sōngdào's Diary*]

*Transcription*：同治十三年十一月一日⋯日中当有黑子，未及见也。

*Translation*: On 09 December 1874 ... In the Sun, there should have been a black spot, while I had not seen it yet.

These transcriptions and translations show that the former two are literally identical, while the third seems to be hearsay evidence. While two of these records share the common terminology, the local treatises (the *Shíménxiànzhì* and the *Zhāngyànzhì*) generally describe the local topics in their region, unless otherwise noted. In this case, their observational sites are considered to be *Zhāngyàn* (張堰: 30°48′N, 121°17′E) and *Shíménxiàn* (石門縣: 30°46′N, 120°45′E). The observational sites for the first two are situated in the suburban area of *Shànghǎi* (上海: 31°14′N, 121°29′E). Considering their geographical proximity and virtually the same text, we need to note the possibility that they may have the same origin, as also shown in the case studies for the eclipse records (Ma, 2004; Lu, 2004). As *Shíménxiànzhì* and *Zhāngyànzhì* are dated as 1879/80 and 1920, respectively, it is probably more plausible to consider that the report in *Shíménxiànzhì* has been copied to *Zhāngyànzhì*. Nevertheless, further surveys would be required for their relationship.

Likewise, the diary records are generally based on the experience and hearsay of the author, unless otherwise noted (the *Guō Sōngdào Rìjì*). While *Guō Sōngdào* was likely to have been in *Xiāngyīn* (湘陰: 28°41′N, 112°53′E) at that time, the record itself is probably based on hearsay. While *Guō Sōngdào* himself probably had not seen this "black spot", his record of hearsay on 09 December1874 lets us confirm the observational date is more likely to be on 09 December, as he has given an entry for every date.

## 4. Observation of the Transit of Venus on 09 December 1874 in East Asia

On the same day, Western astronomical teams were in China and Japan for the observations of the transit of Venus. In China, there were some Western scholars engaging in observations of the transit of Venus in 1874, as reviewed by Lu and Li (2013). Astronomers from the United States, France, and Russia made observations at *Běijīng* (北京: 39°54′N, 116°26′E). They reported its start at 09h 33m local mean time (hereafter, LMT) and its end at 14h 17m LMT (PM, v.28, f.26a) with a drawing of the transit of Venus made by James Craig Watson (華徳孫, 1838 – 1880), as reproduced in Figure 1 (PM, v.28, ff. 26a – 27a). It is recorded that the astronomers at *Shànghǎi* faced an intermittent cover of dense "cloud vapour (雲氣)" and ended up taking some photographs during breaks in the cloud cover (SB: 807, p. 2; see also Table 1).

In Japan, Francisco Diaz Covarrubias, a Mexican astronomer, provided a detailed report entitled *Viaje de la Comision Astronómica Mexicana al Japon* (VCAMJ) accompanied by the observational records and 14 photographs (VCAMJ, pp. 419 – 431) with mean time (*tiempo medio*) at Nogeyama (*Nogue-no-yama*, 野毛山), as reproduced in Figure 2.

The Mexican team reported that at Nogeyama (*Nogue-no-yama*, 野毛山, 35°27′N, 139°37′E) the first contact was at 23h 04m 07.0s on 08 December, the second contact was at 23h 29m 24.6s, the third contact was at 03h 21m 45.4s on 09 December, and the fourth contact at 03h 47m 55.5s. At Yamate (*Del Bluff*, 山手, 35°26′N, 139°39′E) (VCAMJ, p.224), the first contact was at 23h 03m 59.0s on 08 December, the second contact was at 23h 29m 50.0s, the third contact was at 03h 21m 50.9s on 09 December, and the fourth contact at 03h 48m 04.0s. Here, the observers state that they have used the local mean astronomical time (LMT − 12h) that starts at Noon (12:00) in the LMT (VCAMJ, p. 224). The observations at Nogeyama and Yamate in Yokohama indicate that the Venus transit lasted for approximately 4 hours 44 minutes (see Tables 2 – 3).

## 5. Analyses of the Transit of Venus on 09 December 1874

We have used the ephemeris data of JPL-DE430 with the $\Delta T$ (Terrestrial Time (TT) - Universal Time (UT)) value of −3.2 second (USNO, 2018, p. K8), to compute the positions of Venus and the Sun in relation to the Earth during the transit of Venus on 09 December 1874. We computed the LMT of the first, second, third, and fourth contacts of the transit of Venus at *Zhāngyàn* and *Shíménxiàn*. We also computed them at Nogeyama, Yamate, and *Běijīng*. The variation of observational value and calculated value for each case in LMT is within 60 seconds at Nogeyama and Yamate in Japan and at *Běijīng* in China, possibly caused by the difficulty of determining the timing of the four contacts (see Appendix 2 and Figure 4).

The transit of Venus lasted from 9h 50m 30s to 14h 33m 40s at *Shíménxiàn* in LMT. The elevation angle of the Sun was 24° – 37° at *Shíménxiàn* (see Table 4). Generally, it is easier to detect sunspots with the unaided eye when the Sun is near the horizon (*e.g.* Vaquero and Vazquez, 2009). Therefore, it is difficult to conclude that this transit was seen due to the low elevation angle of the Sun. Rather, the intermittent cover of dense "cloud vapour" at *Shànghǎi* (SB: 8, 7, p. 2) is considered to have played the role of a "filter" for the Chinese observers at *Zhāngyàn* and *Shíménxiàn*, situated near *Shànghǎi*.

**6. Contemporary Sunspot Observations and their Areas**

This transit of Venus indicates that the projected (silhouetted) area of Venus on the solar disk satisfies the visibility criteria for an unaided-eye sunspot observation. The computed diameter of Venus is 1.052′ during the transit in 1874, taking the diameter of Venus to be 12,103.6 km (*e.g.* USNO, 2018) and the distance between the Earth and Venus to be 0.2643 AU as computed from JPL-DE430. The distance from the Earth to the Sun is 0.9847 AU (*e.g.* JPL-DE430) and the diameter of the Sun is 1,392,000 km (*e.g.* USNO, 2017). Thus the apparent diameter of the Sun is 32.49′. This means the apparent projected area of Venus on the solar disk is equal to 1048 millionths of visible solar disk (msd).

The value of the projected area of Venus with its apparent diameter of ≈ 1′ is almost at the limit of resolution of human eyes with their visual acuity 20/20 as a definition (*e.g.* Snellen, 1862). This projected area is also consistent with the lower limit of the projected area of a sunspot visible with the unaided eye, while some filter-like objects such as natural filters like cloud/mist cover or instrumental filters like sunglasses are expected for observation. Note that observers with better eyesight can see smaller sunspots with the unaided eye. Willis, Davda, and Stephenson (1996) quoted the mean threshold umbral and whole-spot areas for the detection of a sunspot near the center of the solar disk as being at least 15″ (0.25′) and 41″ (0.68′), respectively, based on the work of Keller and Friedli (1992). Taking the diameter of the Sun to be 32′, these threshold umbral and whole-spot areas become 61 and 452 msd (see also Schaefer, 1993; Usoskin *et al*., 2015).

Admittedly, this is a mathematical estimate and hence can be influenced by both spot brightness, or contrast, and morphology. The brightness of a sunspot umbra (umbral core brightness) is ≈ 10 – 40 % of the brightness of the solar disk (Mathew *et al*., 2007), while that of the silhouette of Venus is ≈ 0 % due to the complete blocking of the sunlight. The Venus disk is also significantly different from sunspots in morphology, as it does not have any penumbrae or scattered spots within a group (*e.g*. Brandt, Schmidt, and Steinegger, 1990) but has a sharp circular limb, although allowance should be made for the Venusian atmosphere in a more detailed investigation.

The "Ledgers Section" of the *Greenwich Photo-heliographic Results* (1907) gives the whole-spot projected areas of Groups 138, 138*, and 140, the three sunspot groups on the solar disk on 09 December 1874, as 80, 115, and 121 msd, respectively. These sunspot areas were measured on the original photograph provided by Harvard College Observatory. Each of the three whole-spot areas is less than the known threshold area for unaided-eye sunspot observations cited in the previous paragraph (452 msd).

In this case, it is clearly difficult to relate the "black spot" seen in East Asia with any of these three sunspot groups. Moreover, the "Footnotes" to the "Measures Section" of the *Greenwich Photo-heliographic Results* (1907) describe the three sunspot groups as follows: "Group 138, December 7 – 11. A line of very small spots; Group 138*, December 9. A few small spots in a straight stream; Group 140, December 9 – 14. A very scattered group, composed at first of four spots …". Regarding the "aggregated" sunspot-group with its whole-spot areas cited above (80, 115, and 121 msd) (see Figure 3 of Willis, Davda, and Stephenson, 1996), if the total group area were dispersed over many small spots spread over a wide zone of the solar disk, the total area would give a poor indication of visibility. Indeed, most spots in such a group are probably well below the unaided-eye detection limit, and the main spot may account only for a small fraction of the total group area, and thus may also be below the detection threshold. The umbral areas of groups 138 and 138* are zero on 09 December and the aggregated umbral area of Group 140 is just 5 msd. Therefore, the area of any "dense black spot" on the solar disk that can be attributed to sunspots is ≤ 5 msd on that day, which is significantly less than 61 msd. In order to compensate for this limitation, the group area could be combined with a morphological classification of the group, if this classification gives an indication of the compactness or dispersion of sunspots within the group in question.

We have plotted in Figure 3 the regions of visibility on the surface of the Earth, of the first contact, the least angular distance, and the fourth contact of Venus with the Sun on 09 December 1874. Figure 3 shows that Japan and China are within the visible region of this transit of Venus, while Western Europe, North America, and South America are mostly outside the visible region of this transit. This visible region explains why the RGO photograph does not show Venus in the solar disk. The RGO glass plates/contact prints on 09 and 12 December 1874 were supplied by Harvard College Observatory (see Willis *et al*., 2013) and the transit of Venus was not visible at Harvard on 09 December 1874.

Therefore, it can be concluded that the "black spot in the Sun" recorded in Chinese records was not a sunspot but Venus transiting the solar disk. This transit was not visible in the western hemisphere, including Harvard College Observatory, but was visible in the eastern hemisphere in countries such as China and Japan. This may explain why Willis, Davda, and Stephenson (1996) found only a negative association between the "black spot in the Sun" on 09 December 1874 in the Chinese records and the tabulated sunspot areas in the Western sunspot records published by the Royal Observatory, Greenwich for the same day.

**7. Visibility of Venus as a Misinterpreted Sunspot near the Solar Limb**

The foregoing discussion provides some insight into the reliability of Chinese local treatises. A "black spot" was potentially visible on 09 December 1874. On this particular date, however, the "black spot" was not a sunspot but the silhouette of the planet Venus.

Figure 4 shows the calculated path of Venus on the solar disk as seen from *Shíménxiàn*. We calculated the positions of Venus relative to the Sun using JPL-DE430. We draw the lines of heliographic longitude and latitude on the solar disk at intervals of 10°. Using the rotational elements of the Sun adopted by Archinal *et al*. (2011a, 2011b), which were based on those by Carrington (1863), we calculate the position angle $P$ of the rotational axis of the Sun, and the heliographic longitude $L_0$ and latitude $B_0$ of the apparent center of the solar disk at the event (4:07 UT on 09 December 1874, for its central value).

Figures 1 and 4 show that Venus should have been seen near the solar limb. On the basis of the geographical location of observational site, the minimum angular distance of the "black spot" from the center of the solar disk is computed to be $\approx 80\%$ of the Sun's radius. This is around 54° – 58° solar latitude. Considering that sunspots at a solar latitude greater than 30 – 40° are considered to be at high latitude, we can state that the transit of Venus in 1874 was at quite high solar latitude (*e.g.* Hathaway, Wilson, and Reichmann, 2003; Arlt, 2009). The relevant records state "a black spot was *in* the Sun". This terminology is consistent with the graphical evidence of *Tiānyuán Yùlì Xiángyìfù* (天元玉曆祥異賦), which interprets "*rìbàng/rìpáng* (日傍/日旁)" as implying not in high solar latitude but *out* of the Sun (Strom, 2015; Hayakawa *et al*., 2017a). This means the local Chinese observers had the capability of observing sunspots with significant area (> 1000 msd) even at high solar latitude and near the solar limb.

This case report also shows the meteorological contribution to the visibility of "black spot(s)" in the solar disk in Chinese observations. While the part of the solar disk at low elevation angles apparently favors unaided-eye sunspot observations (*e.g.* Stephenson and Willis, 1999), this case shows that the intermittent cover of dense "cloud vapor" at *Shànghăi* (SB: 8, 7, p. 2) appeared to make this "black spot" visible to the unaided eye, despite its relatively higher solar altitude (24° – 37°).

**8. Insights on the Comparison Between Chinese and Western Sunspot Observations**

The results presented in this article enhance the overall positive association in the comparison between unaided-eye sunspot observations in the Chinese historical documents and the Western sunspot photographs (heliograms). In the original investigation, Willis*,* Davda, and Stephenson (1996) noted that on 09 December 1874 there were no sunspot groups on the solar disk large enough to be seen with the unaided eye based on direct comparison. These authors suggested a possible transcription error in the observational date, resulting in the alternative date 12 December 1874, when a relatively large sunspot (823 msd) appeared on the solar disk. The salient point of the present investigation is that the original date could be reinstated if allowance is made for the possible misinterpretation of the transit of Venus on 09 December 1874 as a sunspot observation. Thus, the descriptive text of the Chinese record, originally dated 09 December 1874, could be considered to be "doubly secure", even if the date itself might possibly be ambiguous.

At the same time, it should be noted that this Venus transit is not recorded using any of the technical terms for Venus such as *jīnxīng* (金星) (see Figure 1) but is recorded as a "black spot", which is most frequently related with sunspot records (*e.g.* Wittman and Xu, 1987; Yau and Stephenson, 1988; Xu*,* Pankenier, and Jiang, 2000; Hayakawa *et al*., 2017a).

The wrong usage of the terminology for sunspots to describe the transit of Venus suggests that this record was probably not made by contemporary professional astronomers in the court but by local observers. As pointed out by Willis *et al*. (2005) and Hayakawa *et al*. (2017b), the observers of the Chinese historical records were different in the official histories (正史) and the local treatises (地方志). As documented in Lu and Li (2013), the contemporary officials and intellectuals in China were aware of the transit of Venus. Some of the contemporary diaries mentioned that the official observatory was aware of the Venus transit, while observations were apparently at least partially disturbed due to cloud cover (*e.g. Wēng Tónghé Rìjì*, v.2, p. 1073; *Píxiétáishān Sănrén Rìjì*, pp. 449 – 450). The Chinese court tried to predict the transit of Venus at that time and the sickness of the Emperor was associated with this transit by the contemporary Chinese (*Yuèmàntáng Rìjì*, v. 22, f. 2b). Figure 1 in PM also described Venus correctly as *jinxin* (金星) in Chinese a correct term for Venus, while this was issued by Western media.

Therefore, we need to be very careful when we try to extend the reliability of unaided-eye sunspot observations recorded in local treatises to those recorded in the official histories of China. At the very least, we could not find reports of unaided-eye sunspot observations on the same date as a transit of Venus before 1874, using the known catalogues of sunspot observations in China (Yau and Stephenson, 1988; Hayakawa et al., 2015, 2017a; Tamazawa et al., 2017).

## 9. Conclusions and Outlook

In this short contribution, we have examined a record of a Chinese "unaided-eye sunspot observation" on 09 December 1874 in the context of contemporary reports of the transit of Venus, as well as sunspot drawings and photographs. The present investigation emphasizes that the transit of Venus was not seen in the United Kingdom (e.g. at the Royal Greenwich Observatory) or in North America (e.g. at the Harvard College Observatory), but it was seen in the whole of East Asia. Therefore, it is easily explained why observers at Greenwich or Harvard would have missed the "black spot" and yet Chinese local scholars would have been able to see the "black spot" on that day.

In addition, the sunspots on the solar disk on 09 December 1874 were not sufficiently large to be seen with the unaided eye, as explained in Section 6, and hence Willis, Davda, and Stephenson (1996) considered the possibility of transcription error in the date. However, the original date could be reinstated if allowance is made for the possible misinterpretation of the transit of Venus on 09 December 1874 as a sunspot observation. In either case, the descriptive record appears to be accurate. Nevertheless, while the heliograms on both 09 and 12 December 1874 measured by RGO staff were kindly provided by the Harvard College Observatory, these heliograms have probably been lost, unfortunately.

Nevertheless, we still need further investigations for other "negative" pairs according to the direct comparisons between Chinese records and heliograms in Greenwich. There is another example of the original date in the Chinese record being a day with only very small sunspots, but sunspots large enough to be seen with the unaided eye on an acceptable alternative date; for example, 16 February 1904 with the alternative date 24 February 1904 (Willis, Davda, and Stephenson, 1996).

The two cases in the Chinese records when there were at most only small spots on the solar disk on both the original date and any alternative date, namely 15 February 1900 and 30 January 1911, require further investigation. So far, these dates are at least not entirely spotless, as a sunspot group has been reported each for both 15 February 1900 by Tacchini, Wolfer, Hadden, and Catania (Umberto Mazzarella), Lewitzky, and Broger, and for 30 January 1911 by Guillaume at the Lyon Observatory (e.g. Carrasco et al., 2013; Vaquero et al., 2016; see also MS 130.1 in the Catania Observatory). At least, consulting the contemporary sunspot drawings by Alfred Wolfer in Zürich

Observatory, sunspots were certainly visible in the solar disk on 15 February 1900 (HS1304.2:3031), 16 and 24 February 1904 (HS1304.2:4076; HS1304.2:4081), and 30 January 1911 (HS1304.2:5936), but their area is apparently not as large as usual unaided-eye sunspots, which should have an area larger than 452 msd. Consistently, more than a few contemporary observers had apparently overlooked these groups too (see Vaquero *et al*., 2016). Further investigations are required for these cases.

The present case study also indicates that sunspots or silhouettes of bodies (*e.g.* Venus) with an area as large as about 1000 millionths of the visible solar disk (msd) could easily have been seen by the oriental observers using just the unaided eye, under suitable meteorological conditions. Contemporary Chinese reports also indicate that the astronomical records in local treatises were not necessarily derived from contemporary official astronomers and hence should be considered differently from similar records in the official histories.

**Acknowledgments**

This research was conducted under the support of the Grant-in-Aid from the Japan Society for the Promotion of Science (JSPS), Grant Number JP15K05038 (PI: M. Sôma), and JP15H05816 (PI: S. Yoden), and a Grant-in-Aid for JSPS Research Fellows JP17J06954 (PI: H. Hayakawa), as well as a mission project of RISH in Kyoto University. We thank M. Fujiyama for his advice on measurements of sunspot area, Alessandra S. Giunta for providing detailed information for sunspot observations in Catania, and the anonymous reviewers for numerous helpful and constructive comments, including those for the visibility variation of Venus disk and sunspots in the solar disk, and those on consulting sunspot drawings from Zürich Observatory for late unaided-eye sunspot records. We also acknowledge the Research Institute for Humanities of Kyoto University, Toyo Bunko, the National Diet Library of Japan, Staatsbibliothek zu Berlin, ETH Zürich, and Catania Observatory for letting us consult relevant historical documents and sunspot drawings cited in this article.

**Disclosure of Potential Conflicts of Interest**
The authors declare that they have no conflicts of interest.

**Appendix 1: References for Source Documents**
*Guō Sōngdào Rìjì*: 郭嵩燾『郭嵩燾日記』湖南人民出版社, 1981
HS1304.2:3031: *Sonnenfleckenzeichnung Nr. 3026. Zeichnung von Flecken und Fackelflächen zur*

**Appendix 2: Transit of Venus on 09 December 1874**

Here, we show a comparison of LMT of observations and calculations, assuming TT－UT ＝ －3.2 second (USNO, 2018).

Table 1: Timing of Venus Transit at *Běijīng*.

*Běijīng*　　　　116°26′　　　39°54′
北京

| Contacts | Obs | Cal | O －C |
|---|---|---|---|
|  | h m s | h m s | s |
| 1st | 09 33 | 09 32 46 | ＋14 |
| 4th | 14 17 | 14 17 45 | －45 |

Table 2: Timing of Venus Transit at Nogeyama.

Nogeyama　　　139°37′　　35°27′

野毛山

| Contacts | Obs | Cal | O －C |
|---|---|---|---|
|  | h m s | h m s | s |
| First | 11 04 07.0 | 11 03 23.2 | ＋43.8 |
| Second | 11 29 24.6 | 11 30 22.8 | －58.2 |
| Third | 15 21 45.4 | 15 21 11.6 | ＋33.8 |
| Fourth | 15 47 55.5 | 15 48 20.9 | －25.4 |

Table 3: Timing of Venus Transit at Yamate.

Yamate　　　139°39′　　35°26′

山手

| Contacts | Obs | Cal | O －C |
|---|---|---|---|
|  | h m s | h m s | s |
| First | 11 03 59.0 | 11 03 31.0 | ＋28.0 |
| Second | 11 29 50.0 | 11 30 30.7 | －40.7 |
| Third | 15 21 50.9 | 15 21 19.3 | ＋31.6 |
| Fourth | 15 48 04.0 | 15 48 28.6 | －24.6 |

Table 4: Timing of Venus Transit at *Shímónxiàn*.

*Shímónxiàn*　　120°45′　　30°46′

石門縣

| Contacts | Obs | Cal | O －C |
|---|---|---|---|
|  | h m s | h m s | s |
| First |  | 09 50 30.3 |  |
| Second |  | 10 17 43.0 |  |

| | |
|---|---|
| Third | 14 06 29.8 |
| Fourth | 14 33 39.7 |

All of the O – C time differences at Nogeyama and Yamate have the consistent pattern such that the first and third contacts were observed too late and the second and fourth contacts were too early (see also Figure 4). The O – Cs for the first and fourth contacts may be an indication of the difficulty to detect the first/last slight notch in the solar limb produced by the disk of Venus. It is well known that at the second and third contacts the black-drop effect (optical diffraction) could be seen, but that phenomenon usually gives the opposite effect (*i.e.* observers detect the second contact too late and the third contact too early). The O – Cs for the second and third contacts may be due to the optical irradiation, which makes the Sun look larger and Venus on the solar disk look smaller**.**

A Misinterpreted Sunspot Record in 1874
Hayakawa *et al.*, 2019, *Solar Physics*. doi: 10.1007/s11207-019-1504-9

Figure 1: Drawing of the transit of Venus observed at *Běijīng* by James Craig Watson (PM: v.28, f.27a). Note that this drawing is not very precise. The minimum center-to-limb distance is slightly smaller than the calculated value.

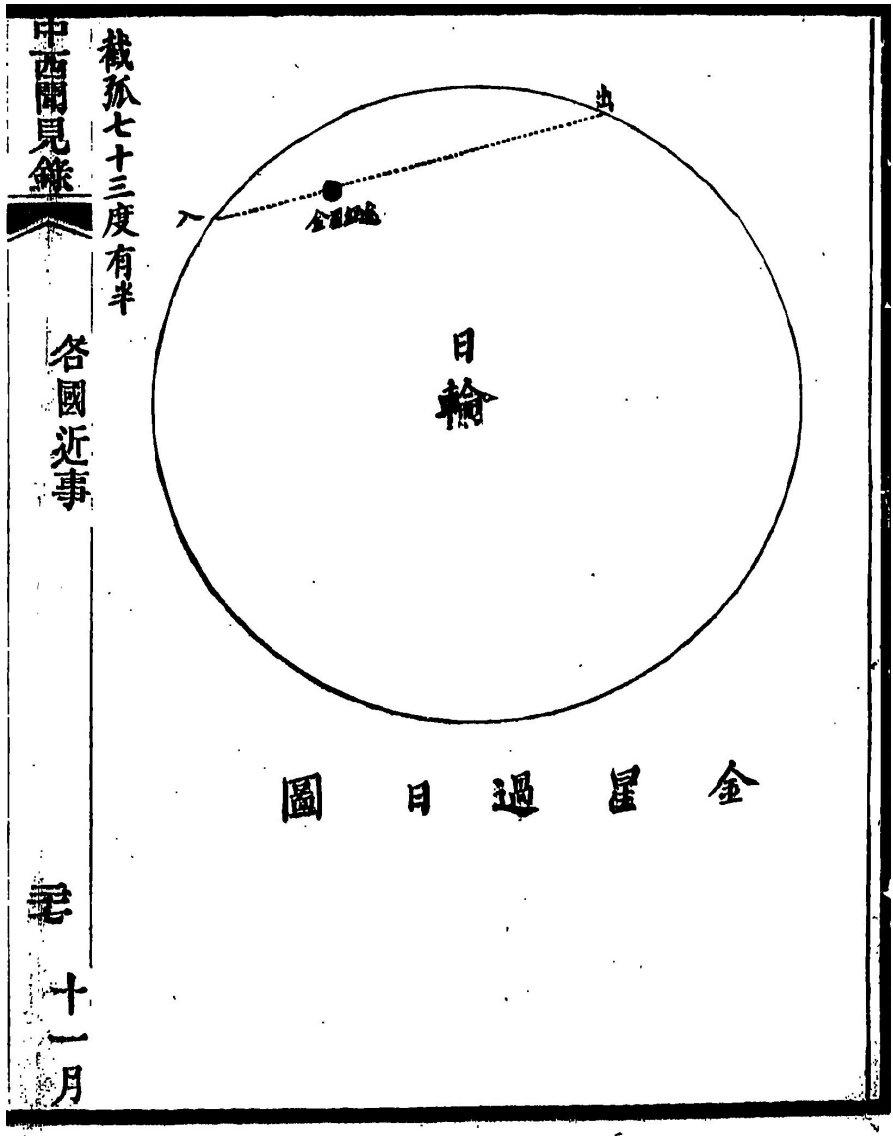

Figure 2: Photographs of Venus transit taken at Nogeyama by Covarrubias (VCAMJ: p.423)

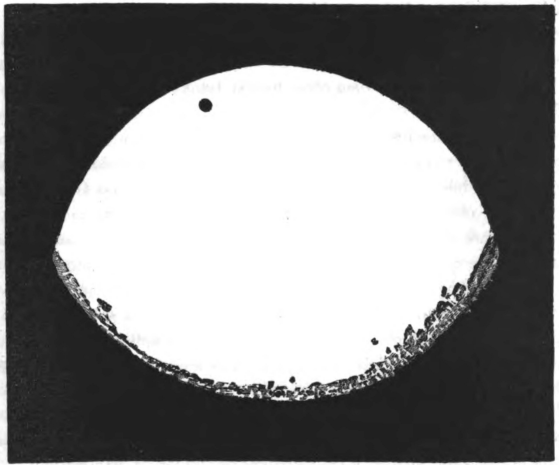

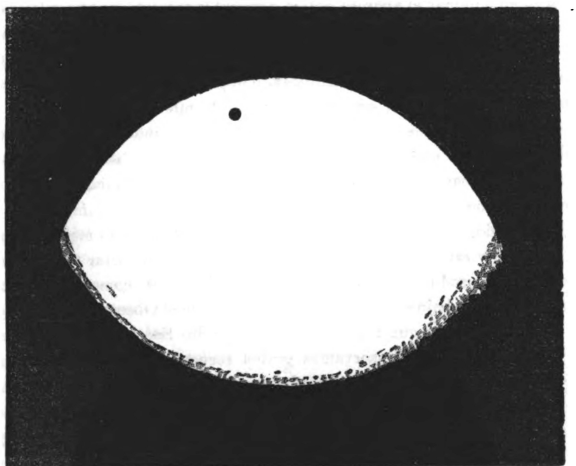

A Misinterpreted Sunspot Record in 1874
Hayakawa *et al.*, 2019, *Solar Physics*. doi: 10.1007/s11207-019-1504-9

Figure 3: Visible zone of the Venus transit in 1874

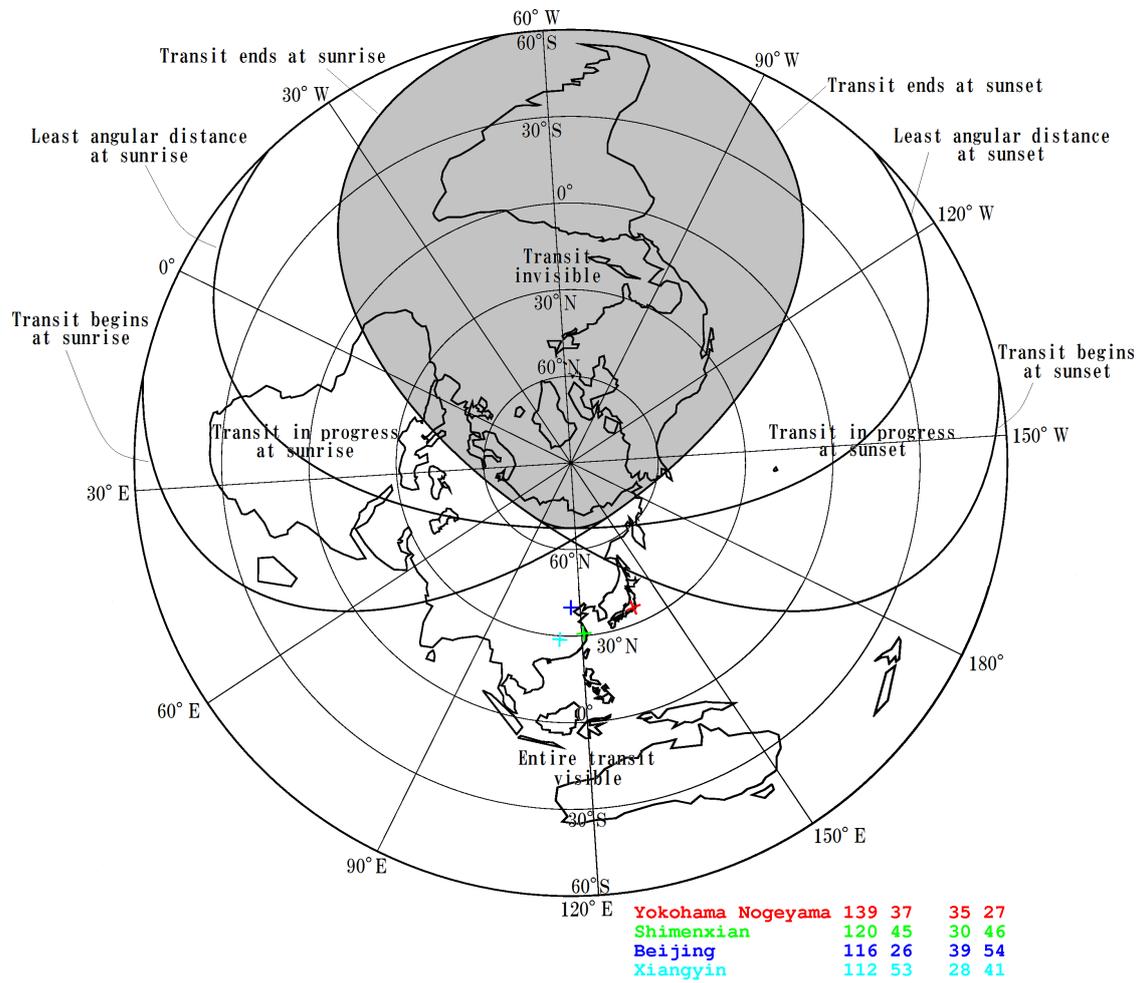

Figure 4: Reconstruction of the transit of Venus in 1874, as observed from *Shíménxiàn*.

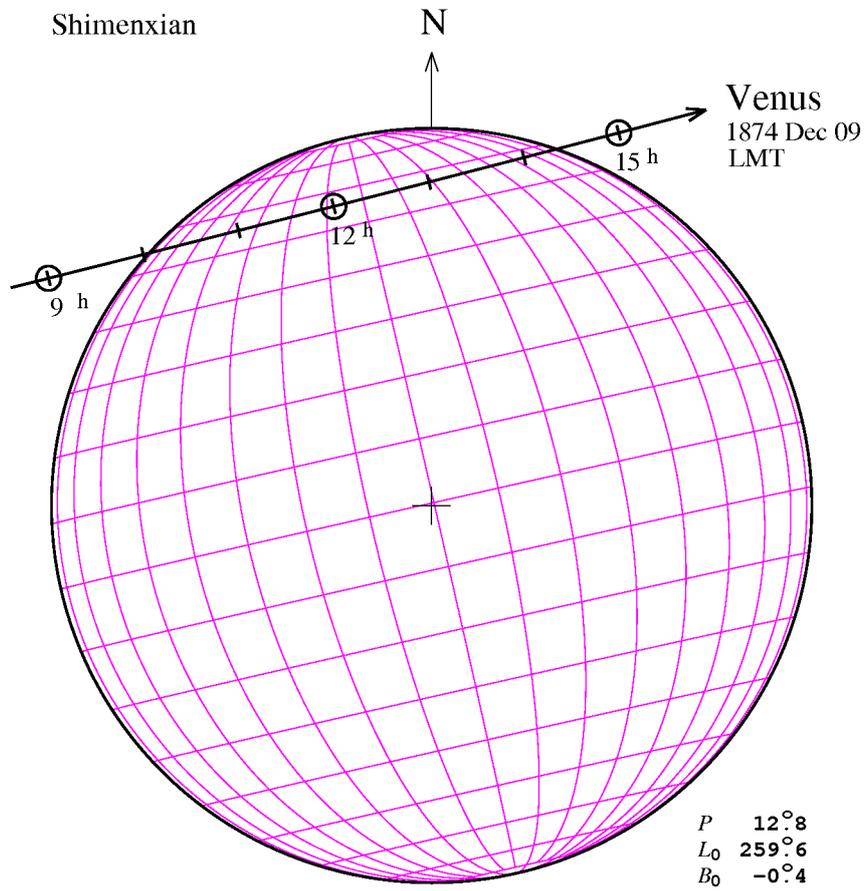